\documentclass[11pt]{article}
\usepackage{amssymb,amsmath,amsfonts}
\usepackage{graphicx}
\usepackage{graphics}
\usepackage{eepic,epsfig}

\begin{document}

\title{Repulsive Casimir-Polder forces from cosmic strings}
\author{A. A. Saharian\thanks{%
E-mail: saharian@ysu.am}, \thinspace\ A. S. Kotanjyan \\
\\
\textit{Department of Physics, Yerevan State University,}\\
\textit{1 Alex Manoogian Street, 0025 Yerevan, Armenia}}
\maketitle

\begin{abstract}
We investigate the Casimir-Polder force acting on a polarizable
microparticle in the geometry of a straight cosmic string. In order to
develop this analysis we evaluate the electromagnetic field Green tensor on
the imaginary frequency axis. The expression for the Casimir-Polder force is
derived in the general case of anisotropic polarizability. In dependence of
the eigenvalues for the polarizability tensor and of the orientation of its
principal axes, the Casimir-Polder force can be either repulsive or
attractive. Moreover, there are situations where the force changes the sign
with separation. We show that for an isotropic polarizability tensor the
force is always repulsive. At large separations between the microparticle
and the string, the force varies inversely as the fifth power of the
distance. In the non-retarded regime, corresponding to separations smaller
than the relevant transition wavelengths, the force decays as the inverse
fourth power of the distance. In the case of anisotropic polarizability, the
dependence of the Casimir-Polder potential on the orientation of the
polarizability tensor principal axes also leads to the moment of force
acting on the particle.
\end{abstract}

\bigskip


\bigskip

\section{Introduction}

As a result of symmetry breaking phase transitions during the evolution of
the early universe various topological defects could arise \cite{Vile85}. In
particular, cosmic strings have been of increasing interest due to the
importance that they may have in cosmology. Though the recent observational
data on the cosmic microwave background radiation have ruled out cosmic
strings as the primary source for primordial density perturbations, they are
still candidates for the generation of a number of interesting physical
effects such as the generation of gravitational waves, high-energy cosmic
rays, and gamma ray bursts. More recently it has been shown that cosmic
stings form in brane inflation models as a by product of the annihilation of
the branes (for a review see \cite{Cope10}).

In the simplest theoretical model the spacetime of a cosmic string is
described by the flat metric with a deficit angle around the string. The
corresponding non-trivial topology results in the distortion of the
zero-point vacuum fluctuations of quantized fields and induces non-zero
vacuum expectation values for physical observables. Another type of vacuum
polarization arises when boundaries are present. The imposed boundary
conditions on quantum fields alter the zero-point oscillations spectrum and
lead to additional shifts in the vacuum expectation values. This is the
well-known Casimir effect (for a review see \cite{Most97}). Combined effects
of topology and boundaries on the quantum vacuum in the geometry of a cosmic
string have been investigated for scalar \cite{Beze06}, vector \cite%
{Brev95,Beze07} and fermionic fields \cite{Aram1,Beze10}, obeying boundary
conditions on cylindrical surfaces (for the Casimir effect in the closely
related geometry of a wedge with cylindrical boundary see Ref. \cite{Nest01}%
). The analysis of the Casimir force for massless scalar fields subject to
Dirichlet, Neumann and hybrid boundary conditions in the setting of the
conical piston has been recently developed in \cite{Fucc11}. The vacuum
polarization effects in a cosmic string spacetime induced by a scalar field
obeying Dirichlet or Neumann boundary conditions on a surface orthogonal to
the string are considered in \cite{Beze11}.

The distortion of the vacuum fluctuations spectrum by the cosmic string also
gives rise to Casimir-Polder forces acting on a polarizable microparticle.
Nowadays, these forces have attracted a great deal of attention because of
their important role in many areas of science \cite{Parseg,Buhm07}. With
advances in nanotechnology and successes in the production of ultracold
atoms it became possible to measure the both Casimir and Casimir-Polder
forces with increased precision (for a review see \cite{Klim09}). In
particular, related to possible applications for micro and nano
electro-mechanical systems, the search for geometries that would lead to
repulsive Casimir-Polder forces is under active research (see, for instance,
discussion in \cite{Milt11} and references therein). In Ref. \cite{Bard10}
the Casimir-Polder force is considered in the static limit for a polarizable
microparticle located near a cosmic string. This approximation is valid at
large distances from the string where the dominant contribution to the force
comes from low frequencies. In the present paper we derive the exact
Casimir-Polder potential for the general case of frequency dependent
anisotropic polarizabilty which allows us to consider the both non-retarded
and retarded regimes. In particular, we show that in the special case of
isotropic polarizability the force is always repulsive.

The paper is organized as follows. In the next section we present our
calculation of the electromagnetic field Green tensor in the frequency
domain for the geometry of a cosmic string. The Casimir-Polder potential and
the force acting on a polarizable microparticle are investigated in sections %
\ref{sec:CP} and \ref{sec:CPF} respectively. The main results are summarized
in section \ref{sec:Conc}. The evaluation of the functions appearing in the
expressions of the components for the Green tensor is given in Appendix.

\section{Electromagnetic field Green tensor}

\label{sec:GT}

We consider the geometry of an infinitely long straight cosmic string. In
the cylindrical coordinates $(x^{1},x^{2},x^{3})=(r,\phi ,z)$ the
corresponding line element has the standard Minkowskian form%
\begin{equation}
ds^{2}=dt^{2}-dr^{2}-r^{2}d\phi ^{2}-dz{}^{2},  \label{ds21}
\end{equation}%
with the difference that now $0\leqslant \phi \leqslant \phi _{0}$ and the
spatial points $(r,\phi ,z)$ and $(r,\phi +\phi _{0},z)$ are to be
identified. The planar angle deficit is related to the mass $\mu _{0}$ per
unit length of the string by $2\pi -\phi _{0}=8\pi G\mu _{0}$, where $G$ is
the gravitational constant. Effective metric with a planar angle deficit
also arises in a number of condensed matter systems (see, for instance, \cite%
{Volo98}). The nontrivial topology due to the cosmic string changes the
structure of the vacuum electromagnetic field. One of consequences is that a
neutral polarizable microparticle placed close to the string experiences a
dispersion force, the Casimir-Polder force. For a microparticle situated at
a point $\mathbf{r}$, the corresponding interaction energy can be expressed
as (see \cite{Buhm07})%
\begin{equation}
U(\mathbf{r})=\frac{1}{2\pi }\int_{0}^{\infty }d\xi \,\,\alpha _{jl}(i\xi
)G_{jl}^{\text{(s)}}(\mathbf{r},\mathbf{r};i\xi ),  \label{CPpot}
\end{equation}%
where $\alpha _{jl}(i\xi )$ is the polarizability tensor of a particle
evaluated at a purely imaginary frequency,
\begin{equation}
G_{jl}^{\text{(s)}}(\mathbf{r},\mathbf{r}^{\prime };\omega )=\int_{-\infty
}^{+\infty }d\tau \,G_{jl}^{\text{(s)}}(x,x^{\prime })e^{i\omega \tau },
\label{Gjls}
\end{equation}%
with $x=(t,\mathbf{r})$, $x^{\prime }=(t^{\prime },\mathbf{r}^{\prime })$, $%
\tau =t-t^{\prime }$, and summation is understood over the indices $%
j,l=1,2,3 $. The expression (\ref{Gjls}) of the Casimir-Polder potential is
given in terms of the subtracted Green tensor%
\begin{equation}
G_{jl}^{\text{(s)}}(x,x^{\prime })=G_{jl}(x,x^{\prime })-G_{jl}^{\text{(M)}%
}(x,x^{\prime }),  \label{Gjls1}
\end{equation}%
where $G_{jl}(x,x^{\prime })$ is the retarded Green tensor for the
electromagnetic field in the geometry of a cosmic string and $G_{jl}^{\text{%
(M)}}(x,x^{\prime })$ is the corresponding tensor in the Minkowski
spacetime. Note that in (\ref{CPpot}) and in what follows we write the
relations in terms of the physical components of the tensors. As the
geometry of a cosmic string is flat outside the string core, the
renormalization procedure is reduced to the subtraction of the Minkowskian
part.

The retarded Green tensor for the electromagnetic field is given by the
expression%
\begin{equation}
G_{jl}(x,x^{\prime })=-i\theta (\tau )\langle E_{j}(x)E_{l}(x^{\prime
})-E_{l}(x^{\prime })E_{j}(x)\rangle ,  \label{Gjl}
\end{equation}%
where $\theta (x)$ is the unit-step function, $E_{j}(x)$ is the operator of
the $j$ component of an electric field, and the angular brackets mean the
vacuum expectation value. By expanding the electric field operator over a
complete set of mode functions $\{E_{\alpha }(x),E_{\alpha }^{\ast }(x)\}$
and using the commutation relations, the following mode sum formula is
obtained for the Green tensor:
\begin{equation}
G_{jl}(x,x^{\prime })=-i\theta (\tau )\sum_{\alpha }\left[ E_{\alpha
j}(x)E_{\alpha l}^{\ast }(x^{\prime })-E_{\alpha l}(x^{\prime })E_{\alpha
j}^{\ast }(x)\right] ,  \label{Gjlms}
\end{equation}%
where asterisk stands for a complex conjugate and the collective index $%
\alpha $ specifies the mode functions.

For the geometry under consideration, we have two different classes of mode
functions corresponding to the cylindrical waves of the transverse magnetic
(TM) and transverse electric (TE) types. They are obtained from the
corresponding mode functions for the vector potential given in Ref. \cite%
{Beze07} and have the form%
\begin{equation}
\mathbf{E}_{\alpha }^{(\lambda )}(x)=\frac{q}{2\pi \gamma \omega }\mathbf{E}%
^{(\lambda )}(r)e^{iqm\phi +ikz-i\omega t},  \label{Ealf}
\end{equation}%
where $m=0,\pm 1,\pm 2,\ldots $, $-\infty <k<+\infty $, $0\leqslant \gamma
<\infty $,%
\begin{equation}
\omega ^{2}=\gamma ^{2}+k^{2},\;q=2\pi /\phi _{0},  \label{omega}
\end{equation}%
and $\lambda =0,1$ correspond to the TM and TE waves, respectively. In (\ref%
{Ealf}), the radial functions for separate components of the electric field,
$E_{l}^{(\lambda )}(r)$, are given by the expressions
\begin{eqnarray}
E_{1}^{(0)}(r) &=&ik\gamma J_{q|m|}^{\prime }(\gamma r),\;E_{2}^{(0)}(r)=-%
\frac{kqm}{r}J_{q|m|}(\gamma r),\;E_{3}^{(0)}(r)=\gamma ^{2}J_{q|m|}(\gamma
r),  \notag \\
E_{1}^{(1)}(r) &=&-\frac{\omega qm}{r}J_{q|m|}(\gamma
r),\;E_{2}^{(1)}(r)=-i\omega \gamma J_{q|m|}^{\prime }(\gamma
r),\;E_{3}^{(1)}(r)=0,  \label{Eir}
\end{eqnarray}%
where $J_{\nu }(x)$ is the Bessel function, the prime means the derivative
with respect to the argument, and the indices $l=1,2,3$ correspond to the
coordinates $r,\phi ,z$, respectively.

Substituting the mode functions (\ref{Ealf}) into the mode sum formula (\ref%
{Gjlms}), the following representation is obtained for the Green tensor:%
\begin{eqnarray}
&& G_{jl}(x,x^{\prime }) =-i\frac{q\theta (\tau )}{2\pi }\sum_{m=-\infty
}^{+\infty }\sum_{\lambda =0,1}\int_{-\infty }^{+\infty }dk\int_{0}^{\infty
}d\gamma \frac{1}{\gamma \omega }  \notag \\
&& \qquad \times \left[ e^{iqm\Delta \phi +ik\Delta z-i\omega \tau
}E_{j}^{(\lambda )}(r)E_{l}^{(\lambda )\ast }(r^{\prime })-e^{-iqm\Delta
\phi -ik\Delta z+i\omega \tau }E_{l}^{(\lambda )}(r^{\prime
})E_{j}^{(\lambda )\ast }(r)\right] ,  \label{Gil}
\end{eqnarray}%
where $\Delta \phi =\phi -\phi ^{\prime }$ and $\Delta z=z-z^{\prime }$. By
using this representation, the spectral component of the Green tensor,
appearing in (\ref{CPpot}), is presented in the form:%
\begin{eqnarray}
&& G_{jl}(\mathbf{r},\mathbf{r}^{\prime };i\xi ) =-\frac{q}{2\pi }%
\sum_{m=-\infty }^{+\infty }\sum_{\lambda =0,1}\int_{-\infty }^{+\infty
}dk\int_{0}^{\infty }d\gamma \frac{1}{\gamma \omega }  \notag \\
&&\qquad \times \left[ E_{j}^{(\lambda )}(r)E_{l}^{(\lambda )\ast
}(r^{\prime })\frac{e^{iqm\Delta \phi +ik\Delta z}}{\omega -i\xi }%
+E_{l}^{(\lambda )}(r^{\prime })E_{j}^{(\lambda )\ast }(r)\frac{%
e^{-iqm\Delta \phi -ik\Delta z}}{\omega +i\xi }\right] .  \label{Gilo1}
\end{eqnarray}%
By taking into account Eq. (\ref{Eir}), we find the expressions
\begin{eqnarray}
G_{11}(\mathbf{r},\mathbf{r}^{\prime };i\xi ) &=&\frac{2q}{\pi }\left[
\partial _{\Delta z}^{2}B(r,r^{\prime },\Delta \phi ,\Delta z)+\frac{2}{%
rr^{\prime }}\partial _{\Delta \phi }^{2}A(r,r^{\prime },\Delta \phi ,\Delta
z)\right] ,  \notag \\
G_{22}(\mathbf{r},\mathbf{r}^{\prime };i\xi ) &=&\frac{2q}{\pi }[\partial
_{\Delta z}^{2}B(r,r^{\prime },\Delta \phi ,\Delta z)-2\partial _{r}\partial
_{r^{\prime }}A(r,r^{\prime },\Delta \phi ,\Delta z)],  \label{Gdiag} \\
G_{33}(\mathbf{r},\mathbf{r}^{\prime };i\xi ) &=&\frac{4q}{\pi }(-\partial
_{\Delta z}^{2}+\xi ^{2})A(r,r^{\prime },\Delta \phi ,\Delta z),  \notag
\end{eqnarray}%
for diagonal components and the expressions
\begin{eqnarray}
G_{12}(\mathbf{r},\mathbf{r}^{\prime };i\xi ) &=&\frac{4q}{\pi rr^{\prime }}%
\left[ 2\partial _{\xi ^{2}}\partial _{\Delta z}^{2}\partial _{\Delta \phi
}+r^{\prime }\partial _{r^{\prime }}\partial _{\Delta \phi }\right]
A(r,r^{\prime },\Delta \phi ,\Delta z),  \notag \\
G_{13}(\mathbf{r},\mathbf{r}^{\prime };i\xi ) &=&-\frac{4q}{\pi }\partial
_{\Delta z}\partial _{r}A(r,r^{\prime },\Delta \phi ,\Delta z),
\label{Gcomp} \\
G_{23}(\mathbf{r},\mathbf{r}^{\prime };i\xi ) &=&-\frac{4q}{\pi r}\partial
_{\Delta \phi }\partial _{\Delta z}A(r,r^{\prime },\Delta \phi ,\Delta z),
\notag
\end{eqnarray}%
for off-diagonal ones. Here we have defined the functions%
\begin{eqnarray}
A(r,r^{\prime },\Delta \phi ,\Delta z) &=&\sideset{}{'}{\sum}_{m=0}^{\infty
}\cos (qm\Delta \phi )\int_{0}^{\infty }dk\cos (k\Delta z)  \notag \\
&& \times \int_{0}^{\infty }d\gamma \frac{\gamma J_{qm}(\gamma
r)J_{qm}(\gamma r^{\prime })}{\omega ^{2}+\Omega ^{2}},  \label{A} \\
B(r,r^{\prime },\Delta \phi ,\Delta z) &=&\sum_{j=\pm 1}\sideset{}{'}{\sum}%
_{m=0}^{\infty }\cos (qm\Delta \phi )\int_{0}^{\infty }dk\cos (k\Delta z)
\notag \\
&&\times \int_{0}^{\infty }d\gamma \frac{\gamma J_{qm-j}(\gamma r)}{\omega
^{2}+\Omega ^{2}}J_{qm-j}(\gamma r^{\prime }),  \label{B}
\end{eqnarray}%
where the prime on the sign of the sum means that the term $m=0$ should be
taken with the weight 1/2. The remained off-diagonal components of the Green
tensor are obtained from those in (\ref{Gcomp}) by using the relation
\begin{equation}
G_{lj}(\mathbf{r}^{\prime },\mathbf{r};-i\xi )=G_{jl}(\mathbf{r},\mathbf{r}%
^{\prime };i\xi ).  \label{OffDiag}
\end{equation}%
In accordance with Eqs. (\ref{Gdiag}) and (\ref{Gcomp}), the evaluation of
the components for the Green tensor is reduced to that for the functions (%
\ref{A}) and (\ref{B}). The corresponding procedure is described in
Appendix. The expressions which will be used below are given by Eqs. (\ref%
{A2}) and (\ref{B3}).

For the evaluation of the Casimir-Polder potential by using Eq. (\ref{CPpot}%
), we need the components of the subtracted Green tensor (\ref{Gjls1}) in
the coincidence limit: $\mathbf{r}^{\prime }\rightarrow \mathbf{r}$. In this
limit the Green tensor $G_{jl}(\mathbf{r},\mathbf{r}^{\prime };i\xi )$ is
divergent. In the problem under consideration, the renormalization is
reduced to the subtraction of the part corresponding to the Green tensor in
the Minkowski spacetime. The components of the Green tensor in a cosmic
string geometry are given by (\ref{Gdiag}) and (\ref{Gcomp}) with the
functions $A(r,r^{\prime },\Delta \phi ,\Delta z)$ and $B(r,r^{\prime
},\Delta \phi ,\Delta z)$ from Eqs. (\ref{A2}) and (\ref{B3}). As it has
been emphasized in Appendix, the $k=0$ terms in these expressions coincide
with the corresponding components in the Minkowski spacetime. Hence, the
subtracted Green tensor is obtained omitting the $k=0$ terms in the
expressions for $G_{jl}(\mathbf{r},\mathbf{r}^{\prime };i\xi )$. It can be
seen that in the coincidence limit the corresponding off-diagonal components
vanish. Further, in the evaluation of the coincidence limit for $\partial
_{\Delta \phi }^{2}A(r,r^{\prime },\Delta \phi ,\Delta z)$, it is convenient
to use the relation
\begin{equation}
\lim_{\mathbf{r}^{\prime }\rightarrow \mathbf{r}}\partial _{\Delta \phi
}^{2}A(r,r^{\prime },\Delta \phi ,\Delta z)=-\lim_{\mathbf{r}^{\prime
}\rightarrow \mathbf{r}}\left[ r\partial _{r}(r\partial _{r})-4r^{2}\partial
_{(\Delta z)^{2}}\left( r\partial _{r}+1\right) \right] A(r,r^{\prime
},\Delta \phi ,\Delta z).  \label{rel3}
\end{equation}%
This relation is obtained from the representation (\ref{A1}) for the
function $A(r,r^{\prime },\Delta \phi ,\Delta z)$ on using the properties of
the modified Bessel function. After a straightforward but long calculation
one then obtains the following closed expression for the nonzero components
of the subtracted Green tensor in the coincidence limit (no summation over $%
l $):%
\begin{equation}
G_{ll}^{\text{(s)}}(\mathbf{r},\mathbf{r};i\xi )=2\xi ^{3}\bigg[ %
\sum_{k=1}^{[q/2]}f_{l}(2\xi rs_{k},s_{k})-\frac{q}{\pi }\sin (q\pi
)\int_{0}^{\infty }dy\frac{f_{l}(2\xi r\cosh (y),\cosh (y))}{\cosh
(2qy)-\cos (q\pi )}\bigg] ,  \label{Gjj}
\end{equation}%
where $[q/2]$ means the integer part of $q/2$ and we have introduced the
notation%
\begin{equation}
s_{k}=\sin (\pi k/q).  \label{sk}
\end{equation}%
In (\ref{fju}), the function $f_{l}(u,v)$ is defined as%
\begin{equation}
f_{l}(u,v)=e^{-u}\sum_{p=1}^{3}b_{lp}(v)u^{p-4},  \label{fju}
\end{equation}%
with%
\begin{equation}
b_{lp}(v)=b_{lp}^{(0)}+b_{lp}^{(1)}v^{2},  \label{bplv}
\end{equation}%
and with the coefficients
\begin{equation}
b_{lp}^{(0)}=\left(
\begin{array}{ccc}
1 & 1 & 1 \\
-2 & -2 & 0 \\
1 & 1 & 1%
\end{array}%
\right) ,\;b_{lp}^{(1)}=\left(
\begin{array}{ccc}
1 & 1 & -1 \\
1 & 1 & -1 \\
0 & 0 & 0%
\end{array}%
\right) ,  \label{bp01}
\end{equation}%
where the rows and columns are numbered by $l$ and $p$ respectively. For
integer values of the parameter $q$, formula (\ref{Gjj}) is reduced to
\begin{equation}
G_{ll}^{\text{(s)}}(\mathbf{r},\mathbf{r};i\xi )=\xi
^{3}\sum_{k=1}^{q-1}f_{l}(2\xi rs_{k},s_{k}).  \label{GjjIntq}
\end{equation}%
Note that the first term in the square brackets of (\ref{Gjj}) dominates for
large values of $q$.

\section{Casimir-Polder potential}

\label{sec:CP}

Given the subtracted Green tensor in the coincidence limit of the arguments,
we can evaluate the Casimir-Polder potential by using the formula (\ref%
{CPpot}). By taking into account Eq. (\ref{Gjj}), the potential is expressed
as%
\begin{equation}
U(\mathbf{r})=\frac{r^{-4}}{16\pi }\sum_{l,p=1}^{3}\bigg[ \sum_{k=1}^{[q/2]}%
\frac{b_{lp}(s_{k})}{s_{k}^{4}}h_{lp}(2rs_{k})-\frac{q}{\pi }\sin (q\pi
)\int_{0}^{\infty }dy\frac{h_{lp}(2r\cosh y)}{\cosh (2qy)-\cos (q\pi )}\frac{%
b_{lp}(\cosh y)}{\cosh ^{4}y}\bigg] ,  \label{U1}
\end{equation}%
where we have defined the function%
\begin{equation}
h_{lp}(y)=\int_{0}^{\infty }dx\,\,x^{p-1}e^{-x}\alpha _{ll}(ix/y).
\label{hpl}
\end{equation}%
For integer values of $q$ the expression for the potential takes the form%
\begin{equation}
U(\mathbf{r})=\frac{r^{-4}}{32\pi }\sum_{k=1}^{q-1}\sum_{l,p=1}^{3}\frac{%
b_{lp}(s_{k})}{s_{k}^{4}}h_{lp}(2rs_{k}).  \label{U1qint}
\end{equation}%
The Casimir-Polder potential, given by Eq. (\ref{U1}), depends on the
distance from the string an on the angles determining the orientation of the
principal axes of the polarizabiltiy tensor with respect to the cosmic
string. These angles enter in the formula through the components $\alpha
_{ll}(i\xi )$ of the polarizability tensor in the cylindrical coordinates $%
(r,\phi ,z)$ with the $z$-axis along the string.

In order to make the dependence on the orientation of the microparticle more
explicit, we introduce the Cartesian coordinates $x_{l}^{\prime }=(x^{\prime
},y^{\prime },z^{\prime })$ with the axes directed along the principal axes
of the polarizability tensor. For the polarizability tensor in this
coordinates one has:%
\begin{equation}
\alpha _{lm}^{\prime }(\omega )=\text{diag}(\alpha _{1}(\omega ),\alpha
_{2}(\omega ),\alpha _{3}(\omega )),  \label{alflm}
\end{equation}%
where $\alpha _{l}(\omega )$ are the corresponding eigenvalues. In addition,
we introduce an intermediate Cartesian coordinates $x_{l}^{\prime \prime
}=(x^{\prime \prime },y^{\prime \prime },z^{\prime \prime })$ with the $%
z^{\prime \prime }$-axis along the string and with the particle location at $%
(r,0,0)$. Introducing the coefficients $\beta _{lm}$ in accordance with $%
x_{l}^{\prime \prime }=r\delta _{l1}+\sum_{m=1}^{3}\beta _{lm}x_{m}^{\prime
} $, it can be seen that (no summation over $l$)
\begin{equation}
\alpha _{ll}(\omega )=\sum_{m=1}^{3}\beta _{lm}^{2}\alpha _{m}(\omega ).
\label{alfllom}
\end{equation}%
The coefficients $\beta _{lm}$, with $\sum_{m=1}^{3}\beta _{lm}^{2}=1$, can
be given in terms of the Euler angles determining the orientation of the
coordinate system $x_{l}^{\prime }$ with respect to the system $%
x_{l}^{\prime \prime }$. For an isotropic particle with $\alpha _{m}(\omega
)\equiv \alpha (\omega )$ one has $\alpha _{ll}(\omega )=\alpha (\omega )$.
The general expression for the matrix $\beta _{lm}$ can be found, for
example, in \cite{Korn00}.

Here we shall consider an important special case of (\ref{alflm}) with
\begin{equation}
\alpha _{1}(\omega )=\alpha _{2}(\omega ).  \label{alf1eq2}
\end{equation}%
In this case the Casimir-Polder potential in Eq. (\ref{U1}) has the form $%
\sum_{l,m=1}^{3}a_{lm}\beta _{lm}^{2}$ with $a_{l1}=a_{l2}$. Now, it can be
seen that%
\begin{equation}
\sum_{m=1}^{3}a_{lm}\beta _{lm}^{2}=a_{l1}-\left( a_{l1}-a_{l3}\right) \beta
_{l3}^{2},  \label{alm}
\end{equation}%
where
\begin{equation}
\beta _{13}=\cos \alpha \sin \beta ,\;\beta _{23}=\sin \alpha \sin \beta
,\;\beta _{33}=\cos \beta .\;  \label{bet13}
\end{equation}%
In (\ref{bet13}), $\beta $ is the angle between the principal axis $%
z^{\prime }$ of the polarizability tensor and the axis of the cosmic string,
$\alpha $ is the angle between the intersection line of the planes $%
(x^{\prime },y^{\prime })$ and $(x^{\prime \prime },y^{\prime \prime })$ and
the axis $y^{\prime \prime }$. The potential does not depend on the third
Euler angle. The latter is a consequence of the symmetry (\ref{alf1eq2}).

Let us consider the asymptotic of the general formula (\ref{U1}) at large
distances from the string. By taking into account that for $y\gg 1$ one has%
\begin{equation}
h_{lp}(y)\approx \alpha _{ll}(0)\Gamma (p),  \label{hpas}
\end{equation}%
with $\alpha _{ll}(0)$ being the static polarizability of a particle, the
potential is presented in the form%
\begin{equation}
U(\mathbf{r})\approx \frac{1}{16\pi r^{4}}\sum_{l,p=1}^{3}\Gamma (p)\alpha
_{ll}(0)\left[ b_{lp}^{(0)}g_{4}(q)+b_{lp}^{(1)}g_{2}(q)\right] .
\label{U1Large}
\end{equation}%
Here we have introduce the notation%
\begin{equation}
g_{n}(q)=\sum_{k=1}^{[q/2]}s_{k}^{-n}-\frac{q}{\pi }\sin (q\pi
)\int_{0}^{\infty }dy\frac{\cosh ^{-n}(y)}{\cosh (2qy)-\cos (q\pi )}.
\label{gn}
\end{equation}%
For the functions $g_{n}(q)$ in (\ref{U1Large}) one has the following
expressions
\begin{equation}
g_{2}(q)=\frac{q^{2}-1}{6},\;g_{4}(q)=\frac{\left( q^{2}-1\right) \left(
q^{2}+11\right) }{90}.  \label{g24}
\end{equation}%
By taking into account the expressions (\ref{bp01}) of the coefficients, for
the potential at large distances one gets (note that $\sum_{p=1}^{3}\Gamma
(p)b_{lp}^{(1)}=0$):
\begin{equation}
U(\mathbf{r})\approx \frac{\left( q^{2}-1\right) \left( q^{2}+11\right) }{%
360\pi r^{4}}\left[ \alpha _{11}(0)-\alpha _{22}(0)+\alpha _{33}(0)\right] .
\label{U1Large1}
\end{equation}%
This expression coincides with the result given in \cite{Bard10} (with the
coefficient corrected) for the static limit. In particular, for an isotropic
polarizability the corresponding force is repulsive. In the special case (%
\ref{alf1eq2}), by taking into account Eqs. (\ref{alfllom}) and (\ref{bet13}%
), we find the asymptotic expression
\begin{equation}
U(\mathbf{r})\approx \frac{\left( q^{2}-1\right) \left( q^{2}+11\right) }{%
360\pi r^{4}}\left[ \alpha _{3}(0)+2\left( \alpha _{1}(0)-\alpha
_{3}(0)\right) \sin ^{2}\alpha \sin ^{2}\beta \right] .  \label{U1Large2}
\end{equation}%
If $\alpha _{3}(0)>\alpha _{1}(0)$, with dependence on the orientation of
the principal axes of the polarizability tensor with respect to the cosmic
string, the force corresponding to Eq. (\ref{U1Large2}) can be either
attractive or repulsive.

In the case of isotropic polarizability, $\alpha _{jl}(\omega )=\delta
_{jl}\alpha (\omega )$, the function $h_{lp}(v)$ in general formula (\ref{U1}%
) does not depend on $l$ and we can explicitly sum over $l$. Introducing the
notation%
\begin{equation}
b_{p}(v)=\sum_{l=1}^{3}b_{lp}(v),  \label{bpv}
\end{equation}%
the expression for the Casimir-Polder potential in the isotropic case is
obtained from (\ref{U1}) by the replacement%
\begin{equation}
\sum_{l=1}^{3}b_{lp}(v)h_{lp}(2rv)\rightarrow b_{p}(v)h_{p}(2rv),
\label{Repl}
\end{equation}%
with $h_{p}(y)=\int_{0}^{\infty }dx\,\,x^{p-1}e^{-x}\alpha (ix/y)$. For the
functions $b_{p}(v)$ one has%
\begin{equation}
b_{1}(v)=b_{2}(v)=2v^{2},\;b_{3}(v)=2-2v^{2}.  \label{bl}
\end{equation}

Now we return to the general formula (\ref{U1}) for the Casimir-Polder
potential. For the further transformation of the corresponding expression
the polarizability tensor should be specified. For the functions $\alpha
_{m}(\omega )$ in (\ref{alflm}) we use the anisotropic oscillator model:%
\begin{equation}
\alpha _{m}(i\xi )=\sum_{j}\frac{g_{j}^{(m)}}{\omega _{j}^{(m)2}+\xi ^{2}},
\label{alfll}
\end{equation}%
where $\omega _{j}^{(m)}$ and $g_{j}^{(m)}$ are the oscillator frequencies
and strengths, respectively. This model for the dynamic polarizability works
well over a wide range of separations. Now, by taking into account (\ref%
{alfllom}), the functions in (\ref{hpl}) are presented in the form%
\begin{equation}
h_{lp}(y)=y^{2}\sum_{m=1}^{3}\sum_{j}g_{j}^{(m)}\beta _{lm}^{2}B_{p}(y\omega
_{j}^{(m)}).  \label{hpl1}
\end{equation}%
with the notation%
\begin{equation}
B_{p}(z)=\int_{0}^{\infty }dx\frac{x^{p-1}e^{-x}}{x^{2}+z^{2}}.  \label{Bp}
\end{equation}%
The expression for the integral in (\ref{Bp}) for general $p$ can be found
in \cite{Prud86}. In the cases needed in (\ref{hpl1}) one has
\begin{eqnarray}
B_{1}(z) &=&z^{-1}\left[ \sin (z)\text{ci}(z)-\cos (z)\text{si}(z)\right] ,
\notag \\
B_{2}(z) &=&-\cos (z)\text{ci}(z)-\sin (z)\text{si}(z),  \label{Bp123} \\
B_{3}(z) &=&-z\left[ \sin (z)\text{ci}(z)-\cos (z)\text{si}(z)\right] +1.
\notag
\end{eqnarray}%
The functions in the right-hand side of Eq. (\ref{Bp123}) are defined as%
\begin{equation}
\text{ci}(x)=-\int_{x}^{\infty }\frac{\cos t}{t}dt,\;\text{si}(x)=\text{Si}%
(x)-\pi /2=-\int_{x}^{\infty }\frac{\sin t}{t}dt.  \label{ci}
\end{equation}%
As a result, for the Casimir-Polder potential we find the following
expressions%
\begin{eqnarray}
U(\mathbf{r}) &=&\frac{1}{4\pi r^{2}}\sum_{l,m,p=1}^{3}\sum_{j}g_{j}^{(m)}%
\beta _{lm}^{2}\bigg[ \sum_{k=1}^{[q/2]}\frac{b_{lp}(s_{k})}{s_{k}^{2}}%
B_{p}(2r\omega _{j}^{(m)}s_{k})  \notag \\
&& -\frac{q}{\pi }\sin (q\pi )\int_{0}^{\infty }dy\frac{B_{p}(2r\omega
_{j}^{(m)}\cosh y)}{\cosh (2qy)-\cos (q\pi )}\frac{b_{lp}(\cosh y)}{\cosh
^{2}y}\bigg] ,  \label{U2}
\end{eqnarray}%
where the functions $b_{lp}(v)$ are defined by Eq. (\ref{bplv}). For the
isotropic case the corresponding formula is obtained from (\ref{U2}) by the
replacement
\begin{equation}
\sum_{l,m=1}^{3}\beta _{lm}^{2}b_{lp}(v)B_{p}(2rv\omega
_{j}^{(m)})\rightarrow b_{p}(v)B_{p}(2rv\omega _{j}),  \label{Repl2}
\end{equation}%
with the functions $b_{p}(v)$ given by (\ref{bl}).

The characteristic length scale in the problem at hand is given by the
wavelengths corresponding to oscillator frequencies $\omega _{j}^{(m)}$. At
small distances from the string, $r\ll 1/\omega _{j}^{(m)}$, the dominant
contribution in (\ref{U2}) comes from the term with $p=1$. By taking into
account that for $z\ll 1$ one has $B_{1}(z)\approx \pi /(2z)$, in the
leading order, for the potential one finds
\begin{equation}
U(\mathbf{r})\approx \frac{1}{16r^{3}}\sum_{l,m=1}^{3}\sum_{j}\eta
_{j}^{(m)}\beta _{lm}^{2}[b_{l1}^{(0)}g_{3}(q)+b_{l1}^{(1)}g_{1}(q)],
\label{U2small}
\end{equation}%
where the function $g_{n}(q)$ is defined in Eq. (\ref{gn}),
\begin{equation}
b_{l1}^{(0)}=(1,-2,1),\;b_{l1}^{(1)}=(1,1,0),  \label{bl00}
\end{equation}%
and we have introduced the notation
\begin{equation}
\eta _{j}^{(m)}=g_{j}^{(m)}/\omega _{j}^{(m)}.  \label{etajm}
\end{equation}%
Note that for odd values of $n$ we have no closed expressions for $g_{n}(q)$%
. The dependence of the asymptotic expression (\ref{U2small}) on the
orientation of the polarizability tensor principal axes is simplified in the
special case (\ref{alf1eq2}):%
\begin{equation}
U(\mathbf{r})\approx \frac{1}{16r^{3}}\sum_{j}\left\{ 2\eta
_{j}^{(1)}g_{1}(q)+(\eta _{j}^{(3)}-\eta _{j}^{(1)})[g_{1}(q)\sin ^{2}\beta
+g_{3}(q)(1-3\sin ^{2}\alpha \sin ^{2}\beta )]\right\} .  \label{U2smallSp}
\end{equation}%
With dependence of the orientation and on the values of $\eta _{j}^{(m)}$,
the corresponding force can be either repulsive or attractive. Moreover, we
can have a situation where the Casimir-Polder force in retarded and
non-retarded regimes has opposite signs. In the isotropic case the
asymptotic expression (\ref{U2small}) reduces to%
\begin{equation}
U(\mathbf{r})\approx \frac{g_{1}(q)}{8r^{3}}\sum_{j}\frac{g_{j}}{\omega _{j}}%
,  \label{U2smallIs}
\end{equation}%
and the corresponding force is repulsive.

\section{Casimir-Polder force}

\label{sec:CPF}

Now we consider the Casimir-Polder force, $\mathbf{F}=-\mathbf{\nabla }U$.
This force is perpendicular to the string and is directed along the radial
direction with the unit vector $\mathbf{n}_{r}$: $\mathbf{F}=F_{r}\mathbf{n}%
_{r}$. In the general case of the polarizability tensor, by taking into
account the expression (\ref{U1}) for the potential, one gets%
\begin{eqnarray}
F_{r} &\mathbf{=}&-\frac{r^{-5}}{16\pi }\sum_{p=1}^{4}\bigg[ %
\sum_{k=1}^{[q/2]}\frac{c_{lp}(s_{k})}{s_{k}^{4}}h_{lp}(2rs_{k})  \notag \\
&& -\frac{q}{\pi }\sin (q\pi )\int_{0}^{\infty }dy\frac{h_{lp}(2r\cosh y)}{%
\cosh (2qy)-\cos (q\pi )}\frac{c_{lp}(\cosh y)}{\cosh ^{4}y}\bigg] .
\label{F1n}
\end{eqnarray}%
In this formula we have defined%
\begin{equation}
c_{lp}(v)=c_{lp}^{(0)}+c_{lp}^{(1)}v^{2},  \label{clpv}
\end{equation}%
with the matrices for the coefficients%
\begin{equation}
c_{lp}^{(0)}=\left(
\begin{array}{cccc}
-3 & -3 & -2 & -1 \\
6 & 6 & 2 & 0 \\
-3 & -3 & -2 & -1%
\end{array}%
\right) ,\;c_{lp}^{(1)}=\left(
\begin{array}{cccc}
-3 & -3 & 0 & 1 \\
-3 & -3 & 0 & 1 \\
0 & 0 & 0 & 0%
\end{array}%
\right) .  \label{clp01}
\end{equation}%
The asymptotic expression for the force at large distances is directly
obtained from (\ref{U1Large1}) and the Casimir-Polder force behaves as $%
1/r^{5}$ power law (retarded regime). In dependence of the eigenvalues for
the polarizability tensor and of the orientation of the principal axes, the
Casimir-Polder force can be either repulsive or attractive. In the case of
isotropic polarizability, the corresponding formula is obtained from (\ref%
{F1n}) by the replacement $\sum_{l=1}^{3}c_{lp}(v)h_{lp}(2rv)\rightarrow
c_{p}(v)h_{p}(2rv)$, where $c_{p}(v)=\sum_{l=1}^{3}c_{lp}(v)$. The functions
$c_{p}(v)$ are given by
\begin{equation}
c_{1}(v)=c_{2}(v)=-6v^{2},\;c_{3}(v)=-2,\;c_{4}(v)=-2+2v^{2}.  \label{c01}
\end{equation}%
Note that in the closely related geometry of a perfectly conducting wedge
the Casimir-Polder force has both radial and azimuthal components. In the
static limit and for an isotropic polarizability tensor, this force has been
considered in \cite{Brev98}. In particular, it has been shown that the
radial component corresponds to an attractive force towards the cusp of the
wedge. The most favorable case for Casimir-Polder repulsion, where the atom
is only polarizable in the $z$ direction, has been recently investigated in
\cite{Milt11}. In the geometry of a wedge, the force on the atom for both
retarded and non-retarded regimes is discussed in \cite{Mend08} for integer
values of $\pi /\phi _{0}$ with $\phi _{0}$ being the opening angle of the
wedge.

In the oscillator model, on the base of (\ref{U2}), for the force we find
the following expression%
\begin{eqnarray}
F_{r} &\mathbf{=}&-\frac{1}{4\pi r^{3}}\sum_{l,m=1}^{3}\sum_{p=1}^{4}%
\sum_{j}g_{j}^{(m)}\beta _{lm}^{2}\bigg[ \sum_{k=1}^{[q/2]}\frac{%
c_{lp}(s_{k})}{s_{k}^{2}}B_{p}(2r\omega _{j}^{(m)}s_{k})  \notag \\
&& -\frac{q}{\pi }\sin (q\pi )\int_{0}^{\infty }dy\frac{B_{p}(2r\omega
_{j}^{(m)}\cosh y)}{\cosh (2qy)-\cos (q\pi )}\frac{c_{lp}(\cosh y)}{\cosh
^{2}y}\bigg] ,  \label{F2}
\end{eqnarray}%
where the functions $B_{p}(z)$, $p=1,2,3$, are given by Eq. (\ref{Bp123}) and%
\begin{equation}
B_{4}(z)=-z^{2}B_{2}(z)+1.  \label{B4}
\end{equation}%
At distances from the string smaller than the relevant transition
wavelengths, the corresponding asymptotic is easily obtained from (\ref%
{U2smallIs}) and the force scales as $1/r^{4}$. For integer values of the
parameter $q$, the formulas for the potential and force are obtained from (%
\ref{U2}), (\ref{F1n}), and (\ref{F2}) omitting the integral terms and
making the replacement (see also (\ref{U1qint}))
\begin{equation}
\sum_{k=1}^{[q/2]}\rightarrow \frac{1}{2}\sum_{k=1}^{q-1}.  \label{Repl3}
\end{equation}

\begin{figure}[tbph]
\begin{center}
\epsfig{figure=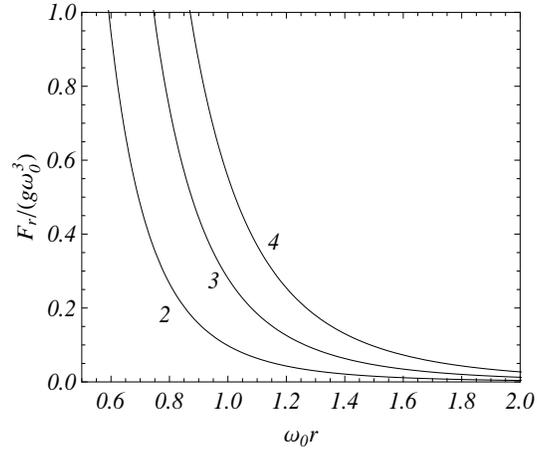, width=7.cm, height=6.cm}
\end{center}
\caption{The Casimir-Polder force as a function of the distance from the
string for separate values of $q$ (numbers near the curves). }
\label{fig1}
\end{figure}

\begin{figure}[tbph]
\begin{center}
\epsfig{figure=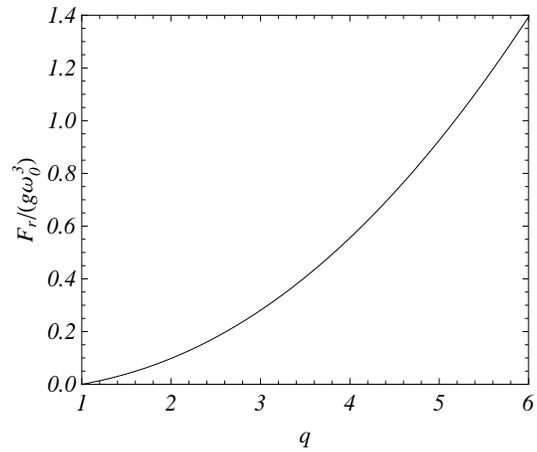, width=7.cm, height=6.cm}
\end{center}
\caption{The Casimir-Polder force as a function of the parameter $q$ for a
fixed separation from the string corresponding to $\protect\omega _{0}r=1$. }
\label{fig2}
\end{figure}

In the case of isotropic polarizability with oscillator frequencies $\omega
_{j}$ and oscillator strengths $g_{j}$, the formula for the Casimir-Polder
force takes the form%
\begin{eqnarray}
F_{r} &\mathbf{=}&-\frac{1}{4\pi r^{3}}\sum_{p=1}^{4}\sum_{j}g_{j}\bigg[ %
\sum_{k=1}^{[q/2]}\frac{c_{p}(s_{k})}{s_{k}^{2}}B_{p}(2r\omega _{j}s_{k})
\notag \\
&& -\frac{q}{\pi }\sin (q\pi )\int_{0}^{\infty }dy\frac{B_{p}(2r\omega
_{j}\cosh y)}{\cosh (2qy)-\cos (q\pi )}\frac{c_{p}(\cosh y)}{\cosh ^{2}y}%
\bigg] ,  \label{F2Iz}
\end{eqnarray}%
with the functions $c_{p}(v)$ defined by Eq. (\ref{c01}). Note that in
accordance with (\ref{c01}) one has $c_{p}(s_{k})<0$ and, hence, the first
term in the square brackets of (\ref{F2Iz}) is always negative. This term
dominates and the Casimir-Polder force in the isotropic case is always
repulsive. In figure \ref{fig1} the Casimir-Polder force (\ref{F2Iz}) is
plotted as a function of the distance for separate values of $q$. The single
oscillator model is used with isotropic polarizability and with the
parameters $g_{j}=g$, $\omega _{j}=\omega _{0}$. Figure \ref{fig2} shows the
Casimir-Polder force in terms of the parameter $q$ for fixed value of the
distance from the string, corresponding to $\omega _{0}r=1$

Note that the dependence of the Casimir-Polder potential on the orientation
of the polarizability tensor principal axes with respect to the string will
also lead to the moment of force acting on the particle. This moment is
obtained differentiating the potential with respect to the respective
angles. As a consequence, the influence of the cosmic string on the system
of particles with anisotropic polarizability results in the macroscopic
polarization.

\section{Conclusion}

\label{sec:Conc}

In the present paper we have investigated the Casimir-Polder force acting on
a polarizable microparticle placed near an infinitely thin straight cosmic
string. As the first step in the evaluation of this force for the general
case of anisotropic polarizability tensor, we have considered the retarded
Green tensor for the electromagnetic field. By using the mode sum formula we
have evaluated all components of this tensor. They are expressed in terms of
two functions (\ref{A}) and (\ref{B}). Employing the summation formula (\ref%
{SumForm}), the representations of these functions are obtained in which the
parts corresponding to the geometry of the Minkowski spacetime are
explicitly extracted. The renormalization of the Green tensor in the
coincidence limit of the spatial arguments is reduced to the omitting of
these parts and the renormalized Green tensor is given by expression (\ref%
{Gjj}).

The investigation of the Casimir-Polder interaction between the
string and a polarizable microparticle, based on the expression
for the Green tensor, is given in sections \ref{sec:CP} and
\ref{sec:CPF}. For the general case of the polarizability
tensor, the Casimir-Polder potential and force are given by the expressions (%
\ref{U1}) and (\ref{F1n}), respectively. The force depends on the distance
from the string and on the angles determining the orientation of the
principal axes of the polarizabiltiy tensor with respect to the cosmic
string. These angles enter in the general formulas through the coefficients $%
\beta _{lm}$ in Eq. (\ref{alfllom}). In dependence of the eigenvalues for
the polarizability tensor and of the orientation of the principal axes, the
Casimir-Polder force can be either repulsive or attractive. Moreover, there
are situations where the force changes the sign with separation. For an
isotropic polarizability tensor the force is always repulsive. At large
distances from the string (retarded regime), the dominant contribution to
the Casimir-Polder potential comes from low frequencies and in the leading
order the potential is given by Eq. (\ref{U1Large1}). The Casimir-Polder
force behaves as $1/r^{5}$ power law. When the two eigenvalues of the
polarizability tensor are equal, the explicit dependence on the orientation
of the particle is given by formula (\ref{U1Large2}). In the special case of
the anisotropic oscillator model for the polarizability tensor, the general
formulas for the Casimir-Polder potential and force are specified to Eqs. (%
\ref{U2}) and (\ref{F2}). In particular, at distances from the string
smaller than the relevant transition wavelengths (non-retarded regime) the
force scales as $1/r^{4}$ power law. In the case of anisotropic
polarizability, the dependence of the Casimir-Polder potential on the
orientation of the polarizability tensor principal axes will also lead to
the moment of force acting on the particle. This results in the macroscopic
polarization induced by the string in the system of polarizable particles.

Generalizing the results of Ref. \cite{Beze11CP}, we can study the
Casimir-Polder interaction for the geometry of a metallic cylindrical shell
with the cosmic string along its axis. The corresponding mode functions are
given in \cite{Beze07}. The results of this investigation will be reported
elsewhere. As it already has been discussed in \cite{Beze07}, from the point
of view of the physics in the region outside the string core, the geometry
with a cylindrical boundary can be viewed as a simplified model for the
superconducting string, in which the string core in what concerns its
superconducting effects is taken to be an ideal conductor.

\section*{Acknowledgments}

A.A.S. was partially supported by PVE/CAPES Program (Brazil). A.A.S.
gratefully acknowledges the hospitality of the Federal University of Para%
\'{\i}ba (Jo\~{a}o Pessoa, Brazil) where part of this work was done.

\appendix

\section{Evaluation of the functions}

\label{sec:App}

In this section, the representations are derived for the functions,
appearing in the expressions (\ref{Gdiag}) and (\ref{Gcomp}) for the
components of the Green tensor. First we consider the function $%
A(r,r^{\prime },\Delta \phi ,\Delta z)$, defined by (\ref{A}). As the first
step we use the relation%
\begin{equation}
\frac{1}{\omega ^{2}+\xi ^{2}}=\int_{0}^{\infty }dx\,e^{-(\omega ^{2}+\xi
^{2})x}.  \label{rel1}
\end{equation}%
Substituting this into (\ref{A}), the integral over $\gamma $ is performed
with the help of the formula \cite{Prud86}%
\begin{equation}
\int_{0}^{\infty }d\gamma \gamma J_{qm}(\gamma r)J_{qm}(\gamma r^{\prime
})\,e^{-\gamma ^{2}x}=\frac{1}{2x}\exp \left( -\frac{r^{2}+r^{\prime 2}}{4x}%
\right) I_{qm}\left( \frac{rr^{\prime }}{2x}\right) .  \label{IntForm1}
\end{equation}%
After integrating over $k$ and changing the integration variable to $%
y=1/(2x) $, we find the following representation%
\begin{eqnarray}
A(r,r^{\prime },\Delta \phi ,\Delta z)&=&\sqrt{\frac{\pi }{8}}%
\int_{0}^{\infty }\frac{dy}{y^{1/2}}\exp \left( -\frac{r^{2}+r^{\prime
2}+(\Delta z)^{2}}{2}y-\frac{\xi ^{2}}{2y}\right)  \notag \\
&& \times \sideset{}{'}{\sum}_{m=0}^{\infty }\cos (qm\Delta \phi
)I_{qm}\left( yrr^{\prime }\right) .  \label{A1}
\end{eqnarray}

For the summation of the series in this expression we use the formula (see
also \cite{Beze11b})%
\begin{equation}
\sideset{}{'}{\sum}_{m=0}^{\infty }\cos (qm\Delta \phi )I_{qm}\left(
w\right) =\frac{1}{2q}\sum_{k}e^{w\cos (2k\pi /q-\Delta \phi )}-\frac{1}{%
4\pi }\sum_{j=\pm 1}\int_{0}^{\infty }dx\frac{\sin (q\pi +jq\Delta \phi
)e^{-w\cosh x}}{\cosh (qx)-\cos (q\pi +jq\Delta \phi )},  \label{SumForm}
\end{equation}%
where in the first term on the right-hand side the summation goes under the
condition%
\begin{equation}
-q/2+q\Delta \phi /(2\pi )\leqslant k\leqslant q/2+q\Delta \phi /(2\pi ).
\label{SumCond}
\end{equation}%
The formula (\ref{SumForm}) is obtained on the base of the integral
representation 9.6.20 from \cite{Abra72} for the modified Bessel function $%
I_{qm}\left( w\right) $. For integer values of $q$, formula (\ref{SumForm})
reduces to the known result \cite{Prud86,Spin08}%
\begin{equation}
\sideset{}{'}{\sum}_{m=0}^{\infty }\cos (qm\Delta \phi )I_{qm}\left(
w\right) =\frac{1}{2q}\sum_{k=0}^{q-1}e^{w\cos (2k\pi /q-\Delta \phi )}.
\label{SumFormSp}
\end{equation}

The application of formula (\ref{SumForm}) to the series in (\ref{A1}),
after the integration over $y$, leads to the following final expression:%
\begin{equation}
A(r,r^{\prime },\Delta \phi ,\Delta z)=\frac{\pi }{4q}\bigg[ \sum_{k}\frac{%
e^{-\xi u_{k}}}{u_{k}}-\frac{q}{2\pi }\sum_{j=\pm 1}\int_{0}^{\infty }dx%
\frac{\sin (q\pi +jq\Delta \phi )e^{-\xi v(x)}/v(x)}{\cosh (qx)-\cos (q\pi
+jq\Delta \phi )}\bigg] .  \label{A2}
\end{equation}%
where we have defined%
\begin{eqnarray}
u_{k} &=&\sqrt{r^{2}+r^{\prime 2}+(\Delta z)^{2}-2rr^{\prime }\cos (2\pi
k/q-\Delta \phi )},  \notag \\
v(x) &=&\sqrt{r^{2}+r^{\prime 2}+(\Delta z)^{2}+2rr^{\prime }\cosh x}.
\label{vx}
\end{eqnarray}

Now we turn to the evaluation of the function $B(r,r^{\prime },\Delta \phi
,\Delta z)$, defined by (\ref{B}). Using (\ref{rel1}), it is presented in
the form%
\begin{eqnarray}
B(r,r^{\prime },\Delta \phi ,\Delta z) &=&\sqrt{\frac{\pi }{8}}%
\int_{0}^{\infty }\frac{dy}{y^{1/2}}\exp \left( -\frac{r^{2}+r^{\prime
2}+(\Delta z)^{2}}{2}y-\frac{\xi ^{2}}{2y}\right)  \notag \\
&&\times \sideset{}{'}{\sum}_{m=0}^{\infty }\cos (qm\Delta \phi )\sum_{j=\pm
1}I_{qm-j}\left( yrr^{\prime }\right) .  \label{B1}
\end{eqnarray}%
By taking into account that%
\begin{equation}
\sum_{j=\pm 1}I_{qm-j}\left( yrr^{\prime }\right) =\frac{2}{yr^{\prime }}%
\partial _{r}I_{qm}\left( yrr^{\prime }\right) ,  \label{rel2}
\end{equation}%
we can rewrite this expression in the form
\begin{eqnarray}
B(r,r^{\prime },\Delta \phi ,\Delta z) &=&\frac{2r}{r^{\prime }}%
A(r,r^{\prime },\Delta \phi ,\Delta z)+\sqrt{\frac{\pi }{2}}\frac{1}{%
r^{\prime }}\partial _{r}\int_{0}^{\infty }\frac{dy}{y^{3/2}}  \notag \\
&&\times \exp \left( -\frac{r^{2}+r^{\prime 2}+(\Delta z)^{2}}{2}y-\frac{\xi
^{2}}{2y}\right) \sideset{}{'}{\sum}_{m=0}^{\infty }\cos (qm\Delta \phi
)I_{qm}\left( yrr^{\prime }\right) .  \label{B2}
\end{eqnarray}%
The second term in the right-hand side is evaluated by using the summation
formula (\ref{SumForm}), in a way similar to that we have used for the
function $A(r,r^{\prime },\Delta \phi ,\Delta z)$. As a result, we find the
following representation:
\begin{eqnarray}
B(r,r^{\prime },\Delta \phi ,\Delta z) &=&\frac{2r}{r^{\prime }}%
A(r,r^{\prime },\Delta \phi ,\Delta z)+\frac{\pi }{2q\xi }\frac{1}{r^{\prime
}}\partial _{r}\bigg[ \sum_{k}e^{-\xi u_{k}}  \notag \\
&& -\frac{q}{2\pi }\sum_{j=\pm 1}\int_{0}^{\infty }dx\frac{\sin (q\pi
+jq\Delta \phi )e^{-\xi v(x)}}{\cosh (qx)-\cos (q\pi +jq\Delta \phi )}\bigg] %
.  \label{B3}
\end{eqnarray}%
The evaluation of the Green tensor components in section \ref{sec:GT} is
based on formulas (\ref{A2}) and (\ref{B3}).

The corresponding expressions in the Minkowski spacetime are obtained taking
$q=1$ in (\ref{A2}) and (\ref{B3}). In this case the integral terms vanish
and we have%
\begin{equation}
A^{\text{(M)}}(r,r^{\prime },\Delta \phi ,\Delta z)=\frac{\pi }{4}\frac{%
e^{-\xi u_{0}}}{u_{0}},\;B^{\text{(M)}}(r,r^{\prime },\Delta \phi ,\Delta z)=%
\frac{\pi }{2}\frac{e^{-\xi u_{0}}}{u_{0}}\cos (\Delta \phi ),  \label{ABM}
\end{equation}%
where $u_{0}$ is given by expression (\ref{vx}) with $k=0$. It is important
to note that the $k=0$ term in the expressions for $qA(r,r^{\prime },\Delta
\phi ,\Delta z)$ and $qB(r,r^{\prime },\Delta \phi ,\Delta z)$ coincide with
$A^{\text{(M)}}(r,r^{\prime },\Delta \phi ,\Delta z)$ and $B^{\text{(M)}%
}(r,r^{\prime },\Delta \phi ,\Delta z)$, respectively. Hence, in the
representations (\ref{A2}) and (\ref{B3}) the Minkowskian parts are
explicitly extracted. As a result, the renormalization procedure is simply
reduced to the omitting of the $k=0$ terms in the corresponding expressions
of the Green tensor components.

\end{document}